\documentclass[cits]{PoS}\usepackage{amsmath} 
\title{Accurate Bottom-Quark Mass from Borel QCD Sum Rules for the
Decay Constants of $B$ and $B_s$ Mesons}\ShortTitle{Accurate
Bottom-Quark Mass from Borel QCD Sum Rules}
\author{Wolfgang Lucha\\Institute for High Energy Physics,
Austrian Academy of Sciences, Nikolsdorfergasse 18, A-1050 Vienna,
Austria\\E-mail: \email{Wolfgang.Lucha@oeaw.ac.at}}
\author{\speaker{Dmitri Melikhov}\\Institute for High Energy
Physics, Austrian Academy of Sciences, Nikolsdorfergasse 18,
A-1050 Vienna, Austria,\\Faculty of Physics, University of Vienna,
Boltzmanngasse 5, A-1090 Vienna, Austria, and\\D.~V.~Skobeltsyn
Institute of Nuclear Physics, Moscow State University, 119991,
Moscow, Russia\\E-mail: \email{dmitri\_melikhov@gmx.de}}
\author{Silvano Simula\\INFN, Sezione di Roma Tre, Via della Vasca
Navale 84, I-00146 Roma, Italy\\E-mail:
\email{simula@roma3.infn.it}}

\abstract{We show that in the context of QCD sum rules a strong
(anti)correlation between the $b$-quark mass $m_b$ and the
$B$-meson's decay constant $f_B$ emerges: $\delta
f_B/f_B\approx-8\,\delta m_b/m_b.$ This observation allows us to
derive a precise value of $m_b$ from a Borel sum rule for the
two-point correlator of heavy--light currents exploiting accurate
$f_B$ results from lattice QCD as input:
$\overline{m}_b(\overline{m}_b)=(4.247\pm0.034)\;\mbox{GeV}.$}

\FullConference{The XXIst International Workshop High Energy
Physics and Quantum Field Theory\\June 23 -- June 30, 2013\\Saint
Petersburg Area, Russia}

\begin{document}\section{Introduction}The standard ``model'' of
elementary particle physics involves, at least, 26 free
parameters~or~28 if neutrinos are not Dirac but Majorana fermions,
most of them related to the fermion-mass sector of the theory. One
of these basic parameters is the mass of the bottom quark. Its
actual numerical~value depends on the choice made for its rigorous
definition; results for this quantity are~usually presented in
terms of either a merely perturbatively given pole mass or, in the
$\overline{\rm MS}$ renormalization scheme, the running mass
$\overline{m}_b(\nu)$ at renormalization scale $\nu$ or the
latter's specific value $m_b\equiv\overline{m}_b(\overline{m}_b)$
at~$\nu=\overline{m}_b.$

In principle, \emph{lattice QCD\/} offers a possibility to infer
the $b$-quark mass from first principles,~\emph{i.e.\/}, directly
from QCD. Unfortunately, the $b$ quark is too heavy for current
lattice setups: some loophole of one kind or the other has to be
found. Moreover, lattice evaluations of the $b$-quark's running
mass involve the calculation of a nonperturbative renormalization
constant; this limits the precision of the mass extraction.
Accordingly, the accuracy of present lattice findings for $m_b$ is
not particularly~high.

Table~\ref{Table:0} summarizes some recent predictions for the
$b$-quark mass found from lattice QCD with unquenched gauge
configurations and two dynamical quarks in the sea by
extrapolating from lighter simulated masses \cite{ETMC1,ETMC2} or
adopting ``heavy-quark effective theory'' (HQET)
\cite{mb_Gimenez,mb_UKQCD,ALPHA} or from moment sum rules for
two-point correlators of \emph{heavy--heavy\/} quark currents that
take advantage of three-loop $O(\alpha_{\rm s}^2)$ \cite{mb4} or
four-loop $O(\alpha_{\rm s}^3)$ \cite{mb1}\footnote{These findings
get support when combining perturbative QCD and lattice QCD with
$2+1$ dynamical sea quarks~\cite{mb2}.} fixed-order
perturbative-QCD results combined with experiment or
renormalization-group-improved
next-to-next-to-leading-logarithmic-order results~plus data
\cite{hoang}.

\begin{table}[h]\small\begin{center}\caption{Bottom-quark mass
$m_b\equiv\overline{m}_b(\overline{m}_b)$ in $\overline{\rm MS}$
renormalization scheme: selection of previous evaluations.}
\label{Table:0}\vspace{2ex}\begin{tabular}{lll}\hline\hline
\multicolumn{1}{c}{Approach}&\multicolumn{1}{c}{Collective of
authors}&\multicolumn{1}{c}{$m_b$ (GeV)}\\\hline Lattice QCD&ETM
Collaboration \cite{ETMC1}&$4.29\pm0.14$\\&ETM Collaboration
\cite{ETMC2}&$4.35\pm0.12$\\&Gimenez {\em et al.}
\cite{mb_Gimenez}&$4.26\pm0.09$\\&UKQCD Collaboration
\cite{mb_UKQCD}&$4.25\pm0.11$\\&ALPHA Collaboration
\cite{ALPHA}&$4.22\pm0.11$\\\hline Moment sum rules&K\"uhn and
Steinhauser \cite{mb4}&$4.191\pm0.051$\\&Chetyrkin {\em et al.}
\cite{mb1}&$4.163\pm0.016$\\&Hoang {\em et al.}
\cite{hoang}&$4.235\pm0.055_{\rm(pert)}\pm0.03_{\rm(exp)}$
\\[.4ex]\hline\hline\end{tabular}\end{center}\end{table}

In the recent study reported here, we used precise values of the
$B_{(s)}$-meson decay constants $f_{B_{(s)}}$ as hadronic input to
\emph{heavy--light\/} Borel QCD sum rules to predict $m_b$ with
comparable accuracy~\cite{lms2013}.

\section{Lesson from Quantum Mechanics: Expect Clear-cut
Anticorrelation of $f_B$ and $m_b$}Our present intention is to
perform a precision determination of the heavy-quark mass
$m_Q=m_b$ from knowledge of the decay constants $f_{B_{(s)}}.$
Within QCD, the question arises: how sensitive are~the numerical
values of these two quantities to each other, what kind and amount
of correlation between them should we expect? To answer this
question, before addressing the real-life problem let us have a
look at the corresponding situation in quantum mechanics. There,
\emph{nonrelativistic potential models\/} are utilized since long
for describing (sufficiently heavy) hadrons as bound states of
quarks~\cite{LSG91,LS99}.\newpage

Now, if the potential involves just one coupling constant, for
instance, if it is a pure Coulomb or pure harmonic-oscillator
potential, for a ground state its wave function at the origin,
$\psi(0),$ is related to its binding energy $\varepsilon$ by
$|\psi(0)|\propto\varepsilon^{3/2};$ for sums of confining and
Coulomb potentials, this relation holds approximately
\cite{lms_qcdvsqm}. Realizing that $|\psi(0)|$ assumes the r\^ole
of a decay constant and exploiting the scaling behaviour of a
heavy-meson decay constant in the heavy-quark limit then relates
the pole mass $m_Q$ of a heavy quark $Q$ to the $B$-meson mass
$M_B,$ approximately by $f_B\,\sqrt{M_B}=\kappa\,(M_B-m_Q)^{3/2}.$
Upon accepting this, it is straightforward to obtain the variation
$\delta f_B$ of $f_B$ as consequence of a small variation $\delta
m_Q$ around some chosen value of $m_Q.$ From the experimental
finding $M_B=5.27\;\mbox{GeV}$ and for $f_B\approx200\;\mbox{MeV}$
near $m_Q\approx4.6\mbox{--}4.7\;\mbox{GeV},$ we get
$\kappa\approx0.9\mbox{--}1.0$ and $\delta f_B\approx-0.5\,\delta
m_Q,$ which~entails$$\frac{\delta f_B}{f_B}\approx-(11\mbox{--}12)
\,\frac{\delta m_Q}{m_Q}\ .$$For instance, $\delta
m_Q=+100\;\mbox{MeV}$ implies $\delta f_B\approx-50\;\mbox{MeV}.$
Hence, we feel entitled to expect a rather high and negative
correlation of $m_b$ and $f_{B_{(s)}}$ manifesting also in QCD
sum-rule predictions~\cite{svz,aliev}.

\section{Earlier Predictions for $B_{(s)}$-Meson Decay Constants
by QCD Sum-Rule Approach}Relying on, essentially, one and the same
expression for the heavy--light correlation function at three-loop
accuracy \cite{chet}, in the last years several QCD sum-rule
extractions of beauty-meson~decay constants have been performed
\cite{nar2001,jamin,lms_fB,nar2012}; their results for $f_B$ are
compiled in Table~\ref{Table:1}. At first glance, all these
findings appear to be consistent and reliable but they are not, as
they do not comply with the quantum-mechanical expectations for
the relationship between $f_B$ and $m_b.$ The crucial issues are
the definition of heavy-quark masses in use and a proper
incorporation of the {\em effective\/} continuum~onset.

\begin{table}[h]\small\begin{center}\caption{$B$-meson decay
constant $f_B$: some predictions by QCD sum rule for heavy--light
two-point function.}\label{Table:1}\vspace{2ex}
\begin{tabular}{lcccc}\hline\hline
&Reference~\cite{nar2001}&Reference~\cite{jamin}
&Reference~\cite{lms_fB}&Reference~\cite{nar2012}\\\hline$m_b$
(GeV)&$4.05\pm0.06$&$4.21\pm0.05$&$4.245\pm0.025$&
$4.236\pm0.069$\\$f_B$ (MeV)&$203\pm23$&$210\pm19$&$193\pm15$&
$206\pm 7$\\\hline\hline\end{tabular}\end{center}\end{table}

After rather successful application \cite{lms_fB,lms_fD} of QCD
sum rules arising from the correlator of two heavy--light
pseudoscalar quark currents to an extraction of the decay
constants of charmed mesons, we recently revisited, \emph{mutatis
mutandis\/} by the same formalism, the beauty-meson system. There,
in contrast to the charmed-meson case, we indeed observe the
presumed \emph{pronounced\/} anticorrelation of heavy-quark mass
and heavy-meson decay constant \cite{lms2013}. Formulating our
correlator in terms of the $\overline{\rm MS}$ running instead of
the pole $b$-quark mass and applying consistent extraction
procedures,~we~find for the QCD-sum rule prediction of $f_B$ a
linear dependence on $m_b$ with negative slope, if keeping the
input values of all other OPE quantities, such as renormalization
scales,~$\alpha_{\rm s},$ quark condensate,~fixed:\begin{equation}
\label{Eq:AC}f_B(m_b)=\left(192.0-37\,\frac{m_b-4.247\;\mbox{GeV}}
{0.1\;\mbox{GeV}}\pm3_{\rm(syst)}\right)\mbox{MeV}\
.\end{equation}This observation suggests to invert, in the
$B_{(s)}$-meson case, our line of reasoning: using, as hadronic
input, our average $f_B^{\rm LQCD}=(191.5\pm7.3)\;\mbox{MeV}$ of
recent lattice-QCD results for $f_B$ \cite{ETMC1,ETMC2,ALPHA,
HPQCD1,HPQCD2,MILC} in our QCD sum rule deriving from the
heavy--light correlator at $O(\alpha_{\rm s}^2)$ accuracy yields
the~accurate estimate $m_b=(4.247\pm0.034)\;\mbox{GeV}.$ In the
following, we present some relevant details of this study.

\section{(Borel-Transformed) QCD Sum Rule from Heavy--Light
Two-Current Correlator}Arising from an evaluation of correlation
functions of appropriate interpolating currents at~both the QCD
level (with quarks and gluons as basic degrees of freedom) and the
hadron level, \emph{QCD sum rules\/} relate the fundamental
parameters of the theory (such as quark masses and strong coupling
$\alpha_{\rm s}$) to experimentally observable features of
hadronic bound states of the QCD degrees of \mbox{freedom. Our}
goal is to adopt this QCD sum-rule approach in order to arrive at
a prediction of the $b$-quark mass $m_b$ from the decay constants
$f_{B_{(s)}}$ of the $B_{(s)}$ mesons. To this end, we start from
the correlator \cite{svz,aliev} of two pseudoscalar currents of a
$b$ quark and a light quark $q$ of mass $m,$ $j_5(x)\equiv(m_b+m)
\,\bar q(x)\,{\rm i}\,\gamma_5\,b(x)$:$$\Pi\!\left(p^2\right)
\equiv{\rm i}\int{\rm d}^4x\exp({\rm i}\,p\,x)\left\langle0\left|
\mbox{T}\!\left(j_5(x)\,j^\dag_5(0)\right)\right|0\right\rangle.$$
At QCD level, Wilson's operator product expansion (OPE)
substitutes nonlocal products of currents by series of local
operators composed of the QCD degrees of freedom, at the price of
introducing --- in addition to perturbative contributions given in
form of integrals of spectral densities $\rho_{\rm pert}(s,\mu)$
--- power corrections of nonperturbative origin, $\Pi_{\rm
power}(\tau,\mu),$ involving so-called vacuum condensates.
Applying to both QCD and hadronic expressions for a correlator
under study a Borel transformation
$\Pi\!\left(p^2\right)\to\Pi(\tau)$ to a Borel variable $\tau$
suppresses at hadron level both higher excitations and hadronic
continuum. The hadronic states above the ground state are subsumed
by integrals of hadron spectral densities $\rho_{\rm hadr}(s)$
with \emph{physical thresholds\/} $s_{\rm phys}$ as lower
endpoints; in our case, $s_{\rm phys}=(M_{B^*}+M_P)^2$ is given by
the beauty vector meson's mass $M_{B^*}$ and the mass $M_P$ of the
lightest pseudoscalar meson with appropriate quantum numbers,
\emph{i.e.\/}, $\pi$ or $K.$ In this way, we get for the QCD sum
rule sought, in terms of the $B_{(s)}$ meson's mass $M_B$ and
decay constant $f_B$ defined by $(m_b+m)\,\langle0|\bar q\,{\rm
i}\,\gamma_5\,b |B\rangle=f_B\,M_B^2,$\begin{align*}
\Pi(\tau)&=f_B^2\,M_B^4\exp\!\left(-M_B^2\,\tau\right)
+\int\limits_{s_{\rm phys}}^\infty\hspace{-.5ex}{\rm d}s
\exp(-s\,\tau)\,\rho_{\rm hadr}(s)\\&=
\int\limits_{(m_b+m)^2}^\infty\hspace{-2.1ex}{\rm d}s
\exp(-s\,\tau)\,\rho_{\rm pert}(s,\mu)+\Pi_{\rm power}(\tau,\mu)\
.\end{align*}\emph{Quark--hadron duality\/} serves to banish all
contributions of higher hadronic states by assuming them to be
counterbalanced by perturbative contributions beyond an
\emph{effective continuum threshold\/} $s_{\rm eff}(\tau)$ that is
an object intrinsic to the QCD sum-rule framework with interesting
and nontrivial facets~\cite{lms_1}, depends on the Borel variable
$\tau$ if requiring rigour in the description of ground-state
properties~\cite{lms_new}, but must not be confused with $s_{\rm
phys}.$ We end up with a QCD sum rule relating~ground state and
OPE:\begin{equation}f_B^2\,M_B^4\exp\!\left(-M_B^2\,\tau\right)
=\int\limits_{(m_b+m)^2}^{s_{\rm eff}(\tau)}\hspace{-2.1ex}{\rm
d}s\exp(-s\,\tau) \,\rho_{\rm pert}(s,\mu)+\Pi_{\rm
power}(\tau,\mu)\ .\label{sr}\end{equation}Even with $\rho_{\rm
pert}(s,\mu)$ and $\Pi_{\rm power}(\tau,\mu)$ known up to a
certain accuracy, the evaluation of this~relation requires us to
formulate both criterion and resulting prescription for
determining the function $s_{\rm eff}(\tau)$ and to assure
reasonable convergence of the OPE. We accomplish the latter by
expanding $\rho_{\rm pert}(s,\mu)$ perturbatively not in terms of
the pole mass \cite{chet} but in terms of the $\overline{\rm MS}$
mass of the $b$ quark. Explicit results for $\rho_{\rm
pert}(s,\mu)$ at three-loop level and $\Pi_{\rm power}(\tau,\mu)$
have been given by Refs.~\cite{chet,jamin}. Table~\ref{Table:2}
presents the numerical values of all OPE quantities adopted as
input to our extraction of $m_b$ \cite{colangelo,pdg}.

\begin{table}[ht]\small\begin{center}\caption{Operator product
expansion inputs: QCD parameters and lowest-dimensional vacuum
condensates.}\label{Table:2}\vspace{2ex}\begin{tabular}{lll}\hline
\hline\multicolumn{1}{c}{OPE quantity}&\multicolumn{1}{c}{Symbol}&
\multicolumn{1}{c}{Numerical input value}\\\hline Light-quark
mass&$\overline{m}_d(2\;\mbox{GeV})$&$(3.5\pm0.5)\;\mbox{MeV}$\\
Strange-quark mass &$\overline{m}_s(2\;\mbox{GeV})$
&$(95\pm5)\;\mbox{MeV}$\\Strong coupling constant&$\alpha_{\rm
s}(M_Z)$&$0.1184\pm0.0007$\\Light-quark condensate&$\langle\bar
qq\rangle(2\;\mbox{GeV})$&$-[(269\pm17)\;\mbox{MeV}]^3$\\
Strange-quark condensate&$\langle\bar ss\rangle(2\;\mbox{GeV})$&
$(0.8\pm0.3)\times \langle\bar qq\rangle(2\;\mbox{GeV})$\\
Two-gluon condensate&$\displaystyle\left\langle\frac{\alpha_{\rm
s}}{\pi}\,GG\right\rangle$&$(0.024\pm0.012)\;\mbox{GeV}^4$\\[1ex]
\hline\hline\end{tabular}\end{center}\end{table}

\section{Effective Continuum Threshold: Allowing for Dependence on
Borel Parameter(s)}Entering in the course of the evaluation of QCD
sum rules at the level of the basic QCD~degrees of freedom, the
effective continuum threshold $s_{\rm eff}$ constitutes,
indisputably, one of the key quantities of the entire formalism:
to a large extent, it determines the numerical value of any hadron
parameter extracted from some QCD sum rule. In order to improve
the output of this QCD sum-rule technique and to acquire, in a
systematic manner, an idea of the \emph{intrinsic\/} uncertainties
of the approach \cite{lms_1}, we collected arguments for a
dependence of this effective continuum threshold on the Borel
parameters introduced, as new variables, into this framework upon
performing Borel transformations \cite{lms_new}, here summarized
by the generic label $\tau$: $s_{\rm eff}=s_{\rm eff}(\tau).$
Surprisingly, the authors of Ref.~\cite{alex} question this $\tau$
dependence; by providing a few clarifying remarks on this issue,
let us try to avoid misconception:\begin{itemize}\item The $\tau$
dependence of the effective continuum threshold is just a trivial
and direct consequence of requiring QCD sum rules such as
Eq.~(\ref{sr}) to be \emph{rigorous\/} relations; from this point
of view, $s_{\rm eff}(\tau)$ is a convenient tool to realize exact
quark--hadron duality and as such \emph{not\/} questionable.\item
Beyond doubt, one may stick to assuming $s_{\rm eff}$ to be a
$\tau$-independent constant. QCD sum~rules of the kind (\ref{sr})
then remain truly \emph{approximate\/} relations; one can then
merely try to minimize the discrepancy between QCD and hadron
sides of one's sum rule in suitably chosen $\tau$ ranges, to
derive in this way some ``best'' $s_{\rm eff}$ value. In actual
extractions, one simultaneously fits both effective continuum
threshold on the QCD side and bound-state features on the hadronic
side.\item Anyway, we should keep in mind one fact: whatever one
does, any bound-state parameter can be extracted from QCD sum
rules only with limited accuracy reflected by its
\emph{systematic\/} error, even if the OPE for the correlator is
known with arbitrarily high accuracy in a limited $\tau$ range,
the Borel window. Thus, in principle \emph{any\/} algorithm for
fixing $s_{\rm eff}$ can be used if it enables one to get a
realistic estimate of this systematic error. Explicit examples
from quantum mechanics (where the ``exact'' bound-state
observables may be found by solving a Schr\"odinger equation) show
that procedures based on $\tau$-independent $s_{\rm eff}$ entail
\emph{uncontrollable\/} errors of the extracted bound-state
properties; we did not succeed in identifying any example where
such a treatment yields a realistic estimate of its systematic
uncertainty \cite{lms_1}. In contrast to this, our procedure,
based on $\tau$-dependent $s_{\rm eff}$ \cite{lms_new}, provides
realistic systematic-error estimates and more precise estimates of
the central values of extracted bound-state parameters compared to
the outcomes if forcing effective continuum thresholds by
arbitrary decision to be $\tau$-independent
constants.\end{itemize}

\section{Reverting the Line of Thought: Calculating the
$\overline{\rm MS}$ Mass $m_b$ of the Bottom Quark}
\begin{figure}[h]\begin{tabular}{cc}
\includegraphics[scale=.3782]{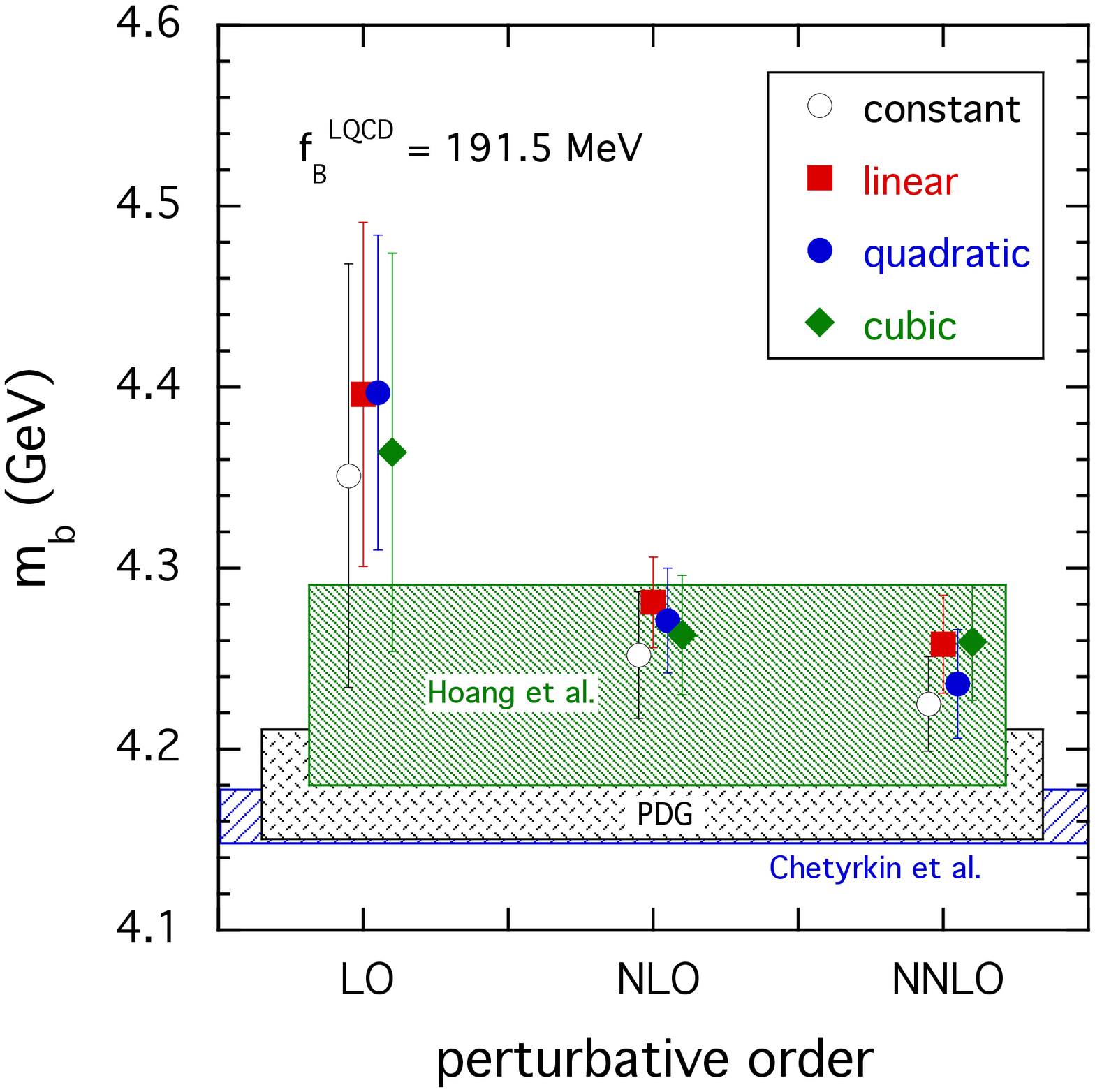}\hspace{1ex}&\hspace{1ex}
\includegraphics[scale=.3782]{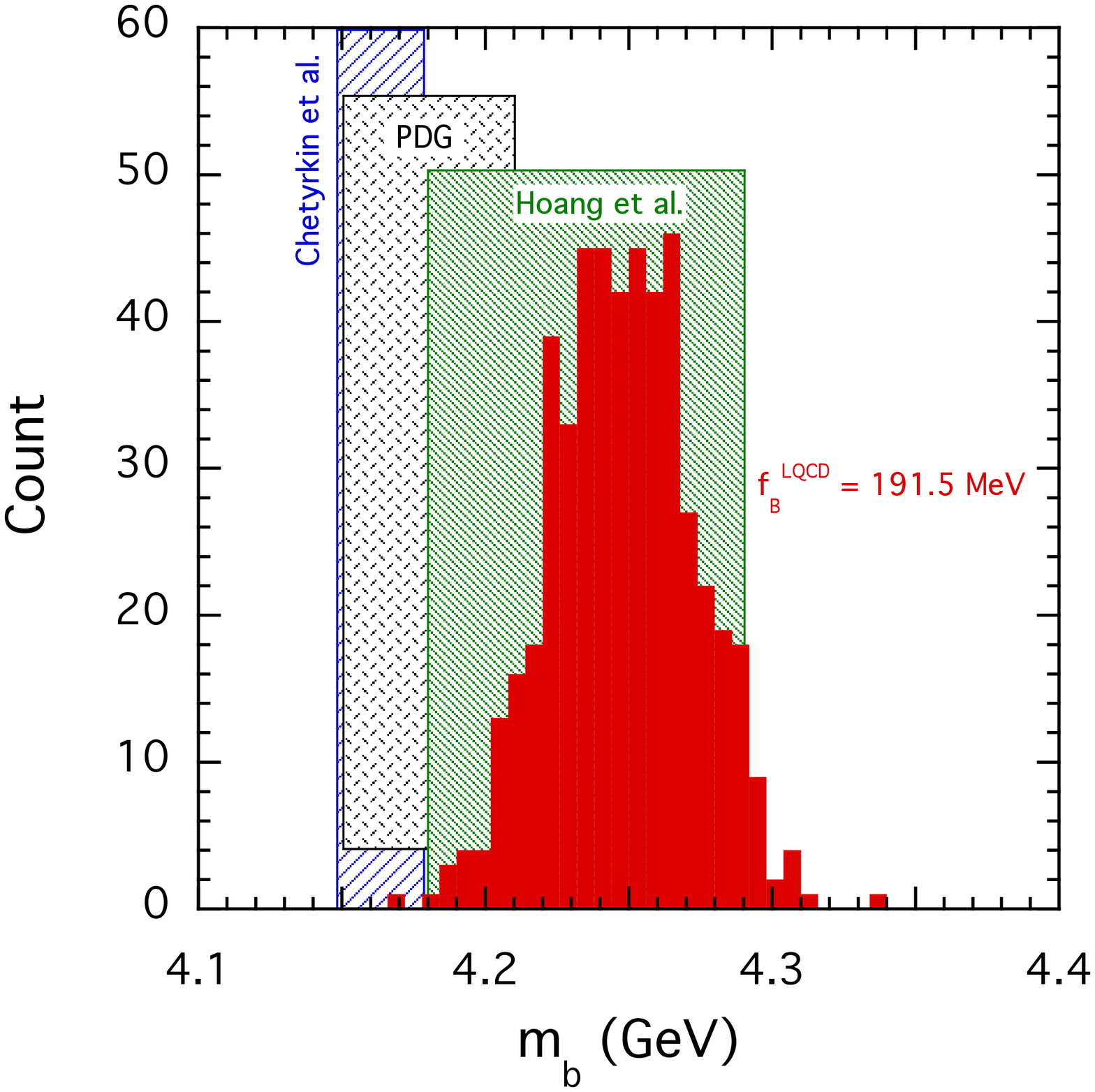}\\[.4574ex](a)&(b)
\end{tabular}\caption{Extraction of the mass of the bottom quark in
$\overline{\rm MS}$ renormalization scheme,
$m_b\equiv\overline{m}_b(\overline{m}_b),$ from our heavy--light
QCD sum rule (\protect\ref{sr}) by a bootstrap analysis of the
errors of all OPE parameters for a central~value of the
beauty-meson decay constant $f_B$ of $f_B=191.5\;\mbox{MeV}$: (a)
Our predictions for $m_b$ calculated for different perturbative
accuracy of the correlator (identified by the labels ``LO,''
``NLO,'' and ``NNLO,'' respectively) and different order of our
polynomial \emph{Ansatz\/} employed for the effective continuum
threshold $s_{\rm eff}(\tau)$ (indicated by ``constant,''
``linear,'' ``quadratic,'' and ``cubic,'' respectively). For
comparison, the ranges corresponding to the ($\pm1\,\sigma$)
errors of the $m_b$ values reported, for instance, by Chetyrkin
{\em et al.} \cite{mb1}, Hoang {\em et al.} \cite{hoang}, and the
Particle Data Group (PDG) \cite{pdg} are represented by the
differently shaded rectangles. (b) Bootstrapping results for the
distribution of masses $m_b$ obtained by assuming Gaussian
distributions for the OPE parameters except for the
renormalization scales $\mu$ and $\nu$ and, for the latter,
uniform distributions in the interval
$3\;\mbox{GeV}<\mu,\nu<6\;\mbox{GeV}.$}\label{Plot:mb}\end{figure}

Even if the
rapid variation (\ref{Eq:AC}) of $f_B$ with $m_b$ renders
difficult to determine $f_B$ from knowledge of $m_b,$ it offers a
possibility to arrive at a precision prediction for $m_b$ by
taking advantage of accurate evaluations of $f_{B_{(s)}}$ provided
by lattice QCD. We seize this opportunity by implementing in the
QCD sum rule (\ref{sr}) the $\tau$ dependence of the effective
continuum threshold $s_{\rm eff}(\tau)$ in form of a polynomial
{\em Ansatz\/} for $s_{\rm eff}(\tau)$ up to third order.
Figure~\ref{Plot:mb} presents a pictorial overview of our
findings. Following the evolution of our $m_b$ results with
increasing perturbative accuracy (\emph{cf}.~Table~\ref{Table:3})
from $O(1)$ leading order (LO) via $O(\alpha_{\rm s})$
next-to-leading order (NLO) to $O(\alpha_{\rm s}^2)$
next-to-next-to-leading order (NNLO), we find for $m_b$ a nice
perturbative convergence, \emph{viz.\/}, a decrease of its central
value and its OPE error.

\begin{table}[h]\small\begin{center}\caption{Bottom-quark mass
$m_b\equiv\overline{m}_b(\overline{m}_b)$ in $\overline{\rm MS}$
renormalization scheme: tracing perturbative convergence.}
\label{Table:3}\vspace{2ex}\begin{tabular}{ll}\hline\hline
\multicolumn{1}{c}{Perturbative order}&\multicolumn{1}{c}{$m_b$
(GeV)}\\\hline Leading order (LO)&$4.38\pm0.1_{\rm(OPE)}
\pm0.020_{\rm(syst)}$\\Next-to-leading order (NLO)&$4.27
\pm0.04_{\rm(OPE)}\pm0.015_{\rm(syst)}$\\Next-to-next-to-leading
order (NNLO)&$4.247\pm0.027_{\rm(OPE)}\pm0.011_{\rm(syst)}$
\\\hline\hline\end{tabular}\end{center}\end{table}

The \emph{OPE uncertainty\/} of our QCD sum-rule extraction of
$m_b$ arises from the uncertainties of the OPE parameters listed
in Table~\ref{Table:2} and from allowing the two renormalization
scales $\mu$ [demanded by the strong coupling $\alpha_{\rm
s}(\mu)$] and $\nu$ [introduced when expressing the $b$-quark pole
mass in terms of the $\overline{\rm MS}$ mass
$\overline{m}_b(\nu)$] to vary independently in the interval
$3\;\mbox{GeV}<\mu,\nu<6\;\mbox{GeV};$ we estimate this error by a
bootstrap analysis. Table~\ref{Table:4} discloses all individual
contributions to our NNLO-level prediction; adding these in
quadrature gives $27\;\mbox{MeV}$ as total OPE error. The
\emph{systematic uncertainty\/} of the QCD sum-rule formalism is
estimated from the spread of results obtained for different {\em
Ans\"atze\/} for~$s_{\rm eff}(\tau).$ Here, it amounts to
$11\;\mbox{MeV}.$ Moreover, the certainly limited accuracy of all
hadronic~input forces us to take into account an additional
uncertainty labelled as \emph{experimental\/}, even if it derives
from~lattice QCD but not from experimental observation. In our
case, $f_B^{\rm LQCD}$ adds a (Gaussian) error of
$18\;\mbox{MeV}.$

\begin{table}[ht]\small\begin{center}\caption{Composition of OPE
uncertainty: contributions by uncertainties of all parameters
entering the~OPE.}\label{Table:4}\vspace{2ex}\begin{tabular}{lr}
\hline\hline \multicolumn{1}{c}{OPE quantity}&\multicolumn{1}{c}
{Individual contribution (MeV)}\\\hline Light-quark mass&4\\Strong
coupling constant&8\\ Quark condensate&20\\ Gluon condensate&7\\
Renormalization scales&14\\\hline\hline\end{tabular}\end{center}
\end{table}

To make a long story short, our findings for the bottom-quark
$\overline{\rm MS}$ mass
$m_b\equiv\overline{m}_b(\overline{m}_b),$ extracted from a Borel
QCD sum rule for the correlator of two heavy--light quark currents
known up to $O(\alpha_{\rm s}^2)$ accuracy by adopting precise
lattice-QCD evaluations of the $B$-meson decay constant as input,
reads\begin{equation}\label{ourmb}m_b=(4.247\pm0.027
_{\rm(OPE)}\pm0.018_{\rm(exp)}\pm0.011_{\rm(syst)})\;\mbox{GeV}\
.\end{equation}Evidently, the systematic error is under control.
Adding all uncertainties in quadrature finally yields
\begin{equation}\label{finres}m_b=(4.247\pm0.034)\;\mbox{GeV}\
.\end{equation}

\section{Summary of Main Results and Conclusions}The observation of
the unexpected scale (\ref{Eq:AC}) of \emph{negative
correlation\/} between $m_b$ and the QCD sum-rule prediction for
$f_B$ forms both basis and starting point of our entire subsequent
investigation:$$\frac{\delta f_B}{f_B}\approx-8\,\frac{\delta
m_b}{m_b}\ .$$Given this behaviour, feeding sufficiently accurate
lattice-QCD values of $f_B$ into our QCD sum-rule machinery
renders possible a precise evaluation of the $b$-quark mass,
culminating in our predictions (\ref{ourmb}) and (\ref{finres})
\cite{lms2013}. Confronted with other published predictions (see
Table~\ref{Table:0}), our $m_b$ result enjoys excellent agreement
with Ref.~\cite{hoang}, acceptable agreement with Ref.~\cite{mb4},
and agreement at the level of two standard deviations with the
Particle Data Group average $m_b=(4.18\pm0.03)\;\mbox{GeV}$
\cite{pdg}; there is, however, undeniable tension with the finding
of Ref.~\cite{mb1} and the value $m_b=(4.171\pm0.009)\;\mbox{GeV}$
by Ref.~\cite{dominguez}. For completeness, with our $m_b$ result
(\ref{finres}) Eq.~(\ref{sr}) predicts, for the $B_{(s)}$
decay~constants,$$f_B=\left(192.0\pm14.3_{\rm(OPE)}
\pm3.0_{\rm(syst)}\right)\mbox{MeV}\ ,\qquad f_{B_s}=\left(228.0
\pm19.4_{\rm(OPE)}\pm4_{\rm(syst)}\right)\mbox{MeV}\ .$$

\vspace{4ex}\noindent{\bf Acknowledgments.} D.M.\ was supported by
the Austrian Science Fund (FWF), project no.~P22843.

\end{document}